\documentstyle[12pt]{article}
\def\la{\mathrel{\mathchoice {\vcenter{\offinterlineskip\halign{\hfil
$\displaystyle##$\hfil\cr<\cr\sim\cr}}}
{\vcenter{\offinterlineskip\halign{\hfil$\textstyle##$\hfil\cr<\cr\sim\cr}}}
{\vcenter{\offinterlineskip\halign{\hfil$\scriptstyle##$\hfil\cr<\cr\sim\cr}}}
{\vcenter{\offinterlineskip\halign{\hfil$\scriptscriptstyle##$\hfil\cr<\cr\sim
\cr}}}}}

\begin{document}

\begin{center}
{\large \bf Unparticle decay of neutrinos and it's effect on 
ultra high energy neutrinos}
\end{center}

\begin{center}
Debasish Majumdar \\
{\it Saha Institute of Nuclear Physics, 1/AF Bidhannagar, 
Kolkata 700 064, India} 
\end{center}
\vskip 3mm
\begin{center}
{\bf Abstract}
\end{center}
{\small 
The unparticle is proposed by Georgi as a conceptual new physics beyond 
the standard model to describe the low energy sector 
of a nontrivial scale invariant sector of an effective theory. We 
consider the neutrino decay in unparticle scenario and investigate 
the effect of such decay for the case of ultra high energy neutrinos 
from GRB. The effect on the detection yield of these neutrinos are 
also probed for a kilometer scale ice-Cerenkov detector. }

\newpage

\section{Introduction} The scale invariance is a feature in which 
a theory or a law 
does not change if the energy scales or length scales are given a 
multiplicative transformation by a common factor. The transformation 
is also known 
as dilatation and this has applications in both mathematics and physics.
In mathematics, the scale invariance can be described by self similarity.
An object is such that if it is magnified by any amount, a smaller piece 
of the object always resembles the whole object. Fractals can be a typical 
example. In statistical mechanics, a scale invariant theory is needed to 
explain the phenomenon of phase transition as the fluctuations near 
the critical point become scale invariant. The scale invariance 
can apply in classical field theory, string theory etc. In  
quantum field theory, the scale invariance, can very naively be 
interpreted as the non-dependendence of the particle interaction strength 
on the energy of the particles. 

In the real world, the scale invariance is manifestly broken by the masses
of the standard model particles. The scale transformation involves mass 
dimension and therefore in scale invariant scenario, the particle masses 
have to be zero. Therefore standard model does not 
have scale invariance. At a scale much higher than that of standard model, 
the scale invariance is restored. 
In a recent work \cite{georgi1,georgi2} Georgi 
observed that if an yet unseen scale invariant sector is present in 
four dimensions and very weakly interacting with the standard model 
particles, then this scale invariant sector contains not ``particles" 
but massless ``Unparticles".  

Under the framework of field theory, an interacting scale invariant 
theory is invariant under conformal group transformation. A conformal 
group transformation is a simple transformation and can be generated 
(in Minkowsky space) by making a scale transformation of the type 
$x^\mu \rightarrow \lambda x^\mu$ ($\mu$ being the space-time coordinates)
\cite{aharony}. Although conformal invariance is well studied in 
two dimensional scenario, it is rare in four dimensional quantum 
field theory. The renormalization group effects break the conformal 
invariance \cite{luo}. In field theory, renormalization group 
flows from some scale invariant ultra violet fixed points to 
some scale invariant infra red fixed points. However, this exceptional  
scenario of conformal invariance in four dimension can be described
by the vector-like non abelian gauge theory with suitable number
of massless fermions. Such a theory is first studied by 
Banks and Zaks (${\cal BZ}$) \cite{bz}. The theory has non-trivial 
infra red fixed points. At low energy limits a non-trivial conformal sector 
is ensured by the presence of non-trivial infra red fixed points in the 
theroy. But this requires 
the number of fermion generations to be non-integral number and therefore 
is not manifested in nature.  

As suggested in Ref. \cite{georgi1,georgi2}  such a nontrivial scale
invariant sector due to ${\cal BZ}$ field might appear at TeV scale.
The ${\cal BZ}$ field interact with the standard model field via 
the exchange of a particle with large mass scale $M_{\cal U}$. This 
can be termed as connector sector \cite{cheung}. Below $M_{\cal U}$ 
standard model and ${\cal BZ}$ field couplings are suppressed by 
powers of  $M_{\cal U}$. Generically the operators can be written 
as \cite{georgi1}, ${\cal O}_{\rm SM} {\cal O}_{\cal BZ} / M^k_{\cal U}$,
($k > 0$) where ${\cal O}_{\rm SM}$ and ${\cal O}_{\cal BZ}$ are 
operators of mass dimensions $d_{\rm SM}$ and $d_{\cal BZ}$ respectively 
and constructed from standard model and ${\cal BZ}$ fields respectively. 
Below the connector sector such interactions between ${\cal BZ}$ and 
standard fields can be parametrized in terms of non-renormilazable 
interactions and the the suppression of nonrenormalizable operators 
by powers of $M^k_{\cal U}$ are induced. As the scale invariance 
sets in at an energy scale $\Lambda_{\cal U}$, the renormalizable 
couplings of the ${\cal BZ}$ fields in the scale invariant sector 
induce dimensional transmutation \cite{coleman} at $\Lambda_{\cal U}$.
The particles in this sector is then described by massless unparticles 
\cite{georgi1}. In the effective theory, below the scale 
$\Lambda_{\cal U}$, the  ${\cal BZ}$ fields match onto the unparticle 
operators and a new set of operators having the form 
$(C_{{\cal O}_{\cal U}} \Lambda^{d_{\cal BZ} - d_{\cal U}} / M^k_{\cal U})
{\cal O}_{\rm SM} {\cal O}_{\cal BZ})\,\,\, ,$
where $d_{\cal U}$ is the scale dimensions of unparticle operator 
${\cal O}_{\cal U}$ and $C_{{\cal O}_{\cal U}}$ is a coefficient to 
be fixed by the matching. This may be occured that the coupling 
between the unparticle and standard model particles may violate 
the conformal invariance but in \cite{georgi2} it is argued that 
the infrared fixed points of the unparticles will not be affected 
by such couplings since ${\cal BZ}$ fields decouple from the standard
model particles at low energies. As observed in Ref. \cite{georgi1},
the unparticle stuff with scale domension $d_{\cal U}$ appears to be 
a nonintegral number $d_{\cal U}$ of massless particles. 

The unparticle physics opens up the possibility of very unexpected 
phenomenology and the effects of unparticle coupling for specific
processes are studied by various authors \cite{unparticles}. Kikuchi
and Okada \cite{kikuchi} adressed the Higgs phenomenology with 
unparticles where they considered operators involving 
scalar unparticle, Higgs and the gauge bosons (gluons and photons) and 
discussed the the effective couplings
between the Higgs boson and the gauge bosons. The interaction 
of  unparticles with Standard Model particles is also addressed 
by Chen and He \cite{chenHe1}. 
Recently Zhou \cite{zhou} discussed  the possibiltiy of neutrino 
decaying into unparticle. Neutrino decay to unparticles are 
also discussed by Chen et al in Ref. \cite{chenHe2}. 
If such neutrino decay really occurs it 
may have consequences in several ongoing neutrino experiments as 
also the future experiments. In a recent work Li et al \cite{Li} has 
discussed the neutrino decay in unparticle scenario as a possible 
explanation of the excess electron like events observed by the 
MiniBooNe \cite{MiniBooNe} neutrino experiment. 
In the present work, the effect of 
such neutrino decay to unparticles is considered on ultra high 
energy (UHE) neutrinos arriving on earth from distant cosmic sources such as 
gamma ray bursts (GRBs). The decay length for decays to unparticles 
depends on quantities like the unparticle-neutrino coupling, the
unparticle scale factor etc. The decay length for this scenario
can be large enough ($\sim$ tens of Mpc) and thus 
can significantly affect the survival probability of a decaying neutrino
if it traverses a baseline length of Mpc order. The ultra high energy 
neutrinos from distant GRBs at astronomical distances from the earth 
may indeed provide such a long baseline. Also, for such a long baseline
the oscillatory effect of neutrinos are averaged out and therefore 
opens up the posibility to effectively probe such decays, if any, in signals 
from UHE neutrinos at terrestrial detectors such as IceCube \cite{icecube}.
    
The paper is organised as follows. In Section 2, the formalism 
is discussed. This includes the neutrino flux from a GRB at 
a redshift $z$; the survival probabilty of such neutrinos 
on reaching the earth if they undergo decay to unparticles.
Therefore the actual flux of neutrinos on reaching the earth 
is obtained from folding the original flux with the
survival probability.
These are described in Section 2.1. In Section 2.2, the analytical 
expressions for the yield of secondary muons and shower events 
induced by the UHE neutrinos at the ice Cerenkov detector are 
described. The calculational procedure of the yield at the 
kilometer scale ice Cerenkov detector such as IceCube for 
the neutrino flux described in Section 2.1, is given in Section 3. 
calculational results are also discussed in this section. Finally,
Section 4 contains discussions and summary. 
 
\section{Formalism}
\subsection{GRB neutrino flux with neutrino decay to unparticles}

The GRB neutrino flux for a particular redshift is estimated 
considering the
the relativistic fireball model \cite{waxman}.
In this type of GRB model, protons (also electrons,
positrons and photons) produced
in the magnetic field of the rotating accretion disc around a possible
black hole are accelerated perpendicular
to the accretion disc at almost the speed of light. This forms a 
jet which is referred to as fireball. The burst is supposed to be
the dissipation of kinetic energy of this relativistic expanding
fireball. High energy pions are photoproduced from $\Delta$ resonance
when the protons in the jet interacts with photons. 
These pions then decay to yield $\nu_\mu$ and
$\nu_e$ in the approximate proportion of 2:1.
                                                                                
The neutrino flux from a GRB depends on several GRB parameters like 
Lorentz boost factor $\Gamma$ (required for the transformation
from the fireball blob to observer's frame of reference), the photon break
energy (as the photon spectrum is considered broken) and the photon luminosity
$L_\gamma$ (generally $\sim 10^{53}$ ergs/sec). 
                                                                                
With all these, the neutrino spectrum from a GRB can be parametrised
as \cite{nayan,pijush}
\begin{equation}
\frac {dN_\nu} {dE_\nu} = A \times {\rm min}(1,E_\nu/E_\nu^b) \times
\frac{1} {E_\nu^2}.
\end{equation}
In the above, $E_\nu$ is the neutrino energy $N_\nu$ is the number of
neutrinos and
\begin{eqnarray}
E_\nu^b &\simeq& 10^6 \frac {\Gamma_{2.5}^2}
{E_{\gamma,{\rm MeV}}^b} {\rm GeV} \nonumber \\
\Gamma_{2.5} &=& \Gamma/10^{2.5} \nonumber \\
A &=& \frac {E_{\rm GRB}} {1 + \ln(E_{\nu{\rm max}}/E_\nu^b)}\,\,,
\end{eqnarray}
where $E_{\nu{\rm max}}$ is the cut-off energy for the GRB neutrinos and
$E_{\rm GRB}$ is the total energy that a GRB emits. Now, the observed
energy $E_\nu^{\rm obs}$ of a neutrino with the actual energy $E_\nu$
coming from a GRB at a redshift distance $z$ is given by the relation
$E_\nu^{\rm obs} = E_\nu/(1+z)$ and similarly, the maximum observable
neutrino energy $E_{\nu{\rm max}}^{\rm obs}$ is
$E_{\nu{\rm max}}^{\rm obs} = E_{\nu{\rm max}}/(1+z)$. The comoving
distance $d$ of a GRB at redshift $z$ is given by
\begin{equation}
d(z) = \frac {c} {H} \int_0^z \frac {dz^\prime}
{\sqrt{\Omega_\Lambda + \Omega_M((1+z^\prime)^3}}
\end{equation}
where $\Omega_M$ is the matter density of the universe at present
epoch,
$\Omega_\Lambda$ is the dark
energy density respectively in units of critical density of
the universe and c,H
are the velocity of light in vacuum and Hubble constant respectively.
In the present calculation $c = 3 \times 10^5$ Km/sec and
$H = 72$ Km/sec/Mpc (1 Mpc $= 3.086 \times 10^{19}$ Km). Therefore the
neutrinos from a single GRB that can be observed on earth per unit
energy per unit area of the earth is given by,
\begin{equation}
\frac {dN_\nu^{\rm obs}} {dE_\nu^{\rm obs}} =
\frac {dN_\nu} {dE_\nu} \frac {1} {4\pi d^2(z)} (1+z)
\end{equation}
                                                                                
The production process of UHE neutrinos suggests that the neutrino
flavours are produced in the ratio $\nu_e : \nu_\mu : \nu_\tau =
1:2:0$. In case of mass flavour oscillation, because of the 
astronomical baseline ($L \sim {\rm Mpc}$),
the acquired relative phases of the propagating neutrino mass eigenstates
is averaged out ($\Delta m^2L/E >> 1$) and the UHE neutrinos from a
GRB reaching the earth are incoherent mixture of mass eigenstates.
In fact, it can be shown \cite{athar} that for   
$\theta_{23} = 45^o$ (maximal mixing) and $\theta_{13}
\simeq 0$, the flavour ratio on reaching the earth, for 
neutrino mass-flavour oscillation with such a long astronomical 
baseline, becomes $\nu_e : \nu_\mu : \nu_\tau = 1:1:1$ irrespective 
of the solar mixing angle. 

However, in the present work we consider neutrino decay to the 
recently proposed ``Unparticle" by Georgi and ivestigate that in the 
event of such decay, the possible signal of UHE neutrinos from distant 
GRBs in terms of muon and shower yields of a kilometer scale detector
such as IceCube. 

We consider here a decay hypothesis where a neutrino 
mass eigen state $\nu_j$ decays to an unparticle and another neutrino 
mass eigenstate $\nu_i$ following the decay relation 
$$
\nu_j \rightarrow {\cal U} + \nu_i
$$
The unparticle physics proposed by Georgi \cite{georgi1} 
indicates a possibility for neutrino decaying into the so called 
unparticle \cite{zhou}.  The effective Lagrangian for the process can be 
written as 
\begin{equation}
{\cal L} = \displaystyle\frac 
{\lambda^{\alpha\beta}_\nu}{\Lambda_{\cal U}^{d_{\cal U}-1}}
            {\bar \nu}_\alpha \gamma_\mu \nu_\beta {\cal O}_{\cal U}
\end{equation}
where ${\cal O}_{\cal U}$ is the unparticle operator, $\lambda_\nu$
is the relevant coupling constant and $\alpha$, $\beta$ are the flavour 
indices. In Eq. (7) dimension transmutation scale is $\Lambda_{\cal U}$ at 
which the scale invariance sets in and $d_{\cal U}$ is the scaling dimension. 

In the mass basis, the interaction term (between neutrino 
and unparticle) can be written as 
\begin{equation}
\displaystyle\frac
{\lambda^{ij}_\nu}{\Lambda_{\cal U}^{d_{\cal U}-1}}
            {\bar \nu}_i \gamma_\mu \nu_j {\cal O}_{\cal U} 
\end{equation}
where $i$, $j$ are the mass eigenstate indices and  
\begin{equation} 
\lambda^{ij}_\nu = \displaystyle\sum_{\alpha,\beta} 
U^*_{\alpha i} \lambda^{\alpha\beta}_\nu U_{\beta j}\,\,\, .
\end{equation} 
In the above $U$, is the Maki-Nakagawa-Sakata (MNS) mixing matrix 
(mass-flavour) for neutrinos. The decay rate for the process 
$\nu_j \rightarrow \nu_i + {\cal U}$ (${\cal U}$ is the unparticle)
can now be elected \cite{zhou} and the neutrino lifetime $\tau$ for such
decay process is given as
\begin{equation}
\displaystyle\frac {\tau_{\cal U}} {m_j} = 
\frac {2^d_{\cal U} \pi^2 d_{\cal U}(2 - d_{\cal U})(d_{\cal U}+1)} 
{3 A_{d_{\cal U}} |\lambda_\nu^{ij}|^2} 
\left ( \frac {\Lambda_{\cal U}^2} {m_j^2} \right )^{d_{\cal U}-1} 
\frac {1} {m_j}\,\,.
\end{equation}
In Eq. (10) above, $m_j$ is the mass of the decaying neutrino 
and $A_{d_{\cal U}}$ is a normalization constant given by \cite{georgi1}
\begin{equation}
A_{d_{\cal U}} = \displaystyle\frac {16\pi^{5/2}} {(2\pi)^{2d_{\cal U}}}
\frac {\Gamma(d_{\cal U} + 1/2)} {\Gamma(d_{\cal U} - 1)\Gamma(2d_{\cal U}) } 
\end{equation}

A decay scenario can be one in which 
both the states $|\nu_2 \rangle$ and $|\nu_3 \rangle$ are unstable 
and decay whereas the lightest state $|\nu_1 \rangle$ is stable. 
Needless to say that the normal mass hierarchy is implicit in this 
scenario. Following 
\cite{pakvasa2}, for this decay scheme with the condition that 
the coherence is lost (baseline $L >> 1/\Delta m^2$ and the oscillatory 
part is absent), the flux of $\nu_a$ for 
flavour $a$ on reaching the earth from a distant GRB at a distance L 
from the earth, is written as 
\begin{equation}
\phi_{\nu_a} (E) = \displaystyle\sum_b\sum_i \phi_{\nu_b}^s (E)
                   |U_{bi}|^2 |U_{ai}|^2 \exp (-4\pi L/(\lambda_d)_i)\,\,.
\end{equation}
In the above, $a$, $b$ are the flavour indices, $i$ is the mass index and
$U$ is the MNS mixing matrix for neutrinos. The exponential term in 
the above equation is the decay term with $(\lambda_d)_i$
is the decay length given by,
\begin{equation}
(\lambda_d)_i = 2.5 km \frac {E} {\rm GeV} \frac {{\rm eV}^2} {\alpha_i}
\end{equation}
where $E$ is the neutrino energy and $\alpha_i (= m_i/\tau)$, $\tau$ being 
the rest frame decay lifetime and $m_i$ is the mass of the $i^{\rm th}$ 
decaying neutrino.
Thus, the decay lifetime in lab frame ($\sim E\tau/m$) and hence the decay 
length $\lambda_d$ has a strong dependence on the neutrino energy too. 

The fluxes of $\nu_e$, $\nu_\mu$ and $\nu_\tau$
from a distant GRB can now be calculated for this decay scenario 
using Eqs (1-4) and Eqs (10-11). Eq. (4) gives the total flux in absence 
of any oscillation or decay. Consdering the fact that at source 
the flux ratio of $\nu_e$, $\nu_\mu$ and $\nu_\tau$ is 1:2:0, 
in absence of decay or oscillation, the $\nu_e$ flux ($\phi_{\nu_e}^s$)
and $\nu_\mu $ flux ($\phi_{\nu_\mu}^s$) and $\nu_\tau$  flux 
($\phi_{\nu_\tau}^s$) can be written in terms of 
$\phi_\nu = \frac {dN_\nu^{\rm obs}} {dE_\nu^{\rm obs}}$,  
\begin{equation}
\phi_{\nu_e}^s = \frac {1} {3}\phi_{\nu}\,,\,\, 
\phi_{\nu_\mu}^s = \frac {2} {3}\phi_{\nu} = 2\phi_{\nu_e}^s  \,,\,\, 
\phi_{\nu_\tau}^s = 0
\end{equation}
Applying the above equation 
for $\nu_e$ flux $\phi_{\nu_e}$ on arrival at earth and with 
the condition that only $|\nu_1 \rangle$ is stable, we get,
\begin{eqnarray}
\phi_{\nu_e} &=& \phi_{\nu_e}^s (E) |U_{e1}|^2 |U_{e1}|^2 + 
                 2\phi_{\nu_e}^s (E) |U_{\mu 1}|^2 |U_{e1}|^2+ \nonumber \\
             && \phi_{\nu_e}^s (E) |U_{e2}|^2 |U_{e2}|^2 
                  \exp (-4\pi L/(\lambda_d)_2)+ \nonumber \\ 
             && 2\phi_{\nu_e}^s (E) |U_{\mu 2}|^2 |U_{e2}|^2 
                  \exp (-4\pi L/(\lambda_d)_2)+  \nonumber \\
             && \phi_{\nu_e}^s (E) |U_{e3}|^2 |U_{e3}|^2 
                  \exp (-4\pi L/(\lambda_d)_3+ \nonumber \\
             && 2\phi_{\nu_e}^s (E) |U_{\mu 3}|^2 |U_{e3}|^2 
                  \exp (-4\pi L/(\lambda_d)_3) \,\,.
\end{eqnarray}
The first two terms on RHS of Eq. (13) (the terms without decay factor)
can be written as 
\begin{equation}
\phi_{\nu_e}^s |U_{e1}|^2 [ |U_{e1}|^2 + 2 |U_{\mu 1}|^2 ]
= |U_{e1}|^2 [ 1 + |U_{\mu 1}|^2 - |U_{\tau 1}|^2 ]
\end{equation}
where the use has been made of the unitarity condition 
$\sum_i U_{ai} U_{bi} = \delta_{ab}$. For $\theta_{13} = 0$ 
($U_{e3}\simeq 0$), $|U_{\mu j}|^2 - |U_{\tau j}|^2 \simeq 0$. The other 
terms of Eq. (17) can also thus be simplified  for $\theta_{13} \simeq 0$. 
With similar approach for $\nu_\mu$ and $\nu_\tau$ fluxes 
($\phi_{\nu_\mu}$ and $\phi_{\nu_\tau}$ respectively) can be 
calculated. The expressions for three neutrino fluxes on arrival at the 
earth are simplified as (in terms of the source flux) 
\begin{eqnarray}
\phi_{\nu_e} &=& \phi_{\nu_e}^s (E) [ |U_{e 1}|^2 + 
                 |U_{e2}|^2 \exp (-4\pi L/(\lambda_d)_2) \nonumber \\
             &+& |U_{e3}|^2 \exp (-4\pi L/(\lambda_d)_3) \nonumber \\
&& \nonumber \\
\phi_{\nu_\mu} &=& \phi_{\nu_e}^s (E) [ |U_{\mu 1}|^2 + 
                 |U_{\mu 2}|^2 \exp (-4\pi L/(\lambda_d)_2) \nonumber \\
             &+& |U_{\mu 3}|^2 \exp (-4\pi L/(\lambda_d)_3) \nonumber \\
&& \nonumber \\
\phi_{\nu_\tau} &=& \phi_{\nu_e}^s (E) [ |U_{\tau 1}|^2 + 
                 |U_{\tau 2}|^2 \exp (-4\pi L/(\lambda_d)_2) \nonumber \\
             &+& |U_{\tau 3}|^2 \exp (-4\pi L/(\lambda_d)_3)
\end{eqnarray} 

From Eqs. (10,13,15)
one can see that in case of $L >> \lambda_d$, the decay 
effect is washed out. The decay lifetime ($\tau$) becomes so small that 
the neutrino under consideration decays completely much before it 
reaches the earth. Under this condition the flavour ratio  
takes the form $\phi_{\nu_e} : \phi_{\nu_\mu} : \phi_{\nu_\tau} 
= |U_{e1}|^2 : |U_{e2}|^2 : |U_{e3}|^2$ (from Eq. 15) as discussed in 
\cite{pakvasa2}. In this case, the actual distance 
of UHE neutrino sources such as 
GRB is not very important and one can work with the diffuse GRB neutrino
flux such as given in Ref. \cite{raj2}. 
But, if $\lambda_d \sim L$ then one cannot
neglect the exponential term in Eqs. (13) and (15) and knowledge 
of $L$ is essential.
In such decay event, it is useful to work with neutrino fluxes from 
GRBs with definite redshifts ($z$) (and hence different baselines $L$)
in order to probe the effects of neutrino decay. 
In the present work, 
oscillation parameters are chosen as $\theta_{23} = 45^o$, $\theta_{13} = 0$
and $\theta_{12} = 32.31^o$ and redshift $z = 0.03$ unless otherwise 
mentioned.

It may be noted here, that for a GRB with redshift $z = 0.03$ 
($\sim 10^{21}$ km $= L$), the decay length $\lambda_d \simeq 4 \times 10^{20}$ 
km for unparticle coupling $\lambda = 0.004$, $d = 1.3$ and 
$\Lambda_{\cal U} = 1$ TeV at neutrino energy 
$E_\nu \simeq 2$ TeV. Therefore for neutrino decay into an unparticle,
the neutrino decay length is indeed of the order of the baseline length $L$.

\subsection{Detection of UHE neutrinos}

The ultra high energy neutrinos from a GRB can be detected in a
terrestrial detector such as IceCube by detecting the secondary 
products like muons, tauons, electromagnetic or hadronic showers
that are produced due to charged 
current (CC) and neutral current (NC) interactions of UHE neutrinos
with the terrestrial rock and detector material. 
The IceCube is a 1 km$^3$ ice/water Cerenkov detector 
in south pole ice. The secondary muons that are produced via CC
interaction can be detected by track signal. The $\nu_\tau$ CC 
interactions produce the so called ``Double Bang" events (track + shower)
whereas $\nu_e$ CC interactions gives electromagnetic shower at the detector.
In the present calculation however, we do not consider the 
``double bang" events as this is perhaps effective for a narrow 
energy range of 1 PeV to 20 PeV for kilometer cube detector \cite{nayantau}.
For $\nu_\tau$ CC interactions we consider instead  the decay channel 
($\tau \rightarrow \bar{\nu_\mu} \mu \nu_\tau$) of secondary tauons 
where the muons thus produced give track signals. The NC interactions 
of all neutrino flavours produce shower events.

The total number of secondary muons induced by GRB neutrinos at a
detector of unit area is given by (following \cite{gaisser,raj1,nayan})
\begin{equation}
S = \int_{E_{\rm thr}}^{E_{\nu{\rm max}}^{\rm obs}} 
dE_\nu^{\rm obs} \frac {dN_{\nu}^{\rm obs}} {dE_\nu^{\rm obs}} P_{\rm surv}
(E_\nu^{\rm obs},\theta_z) P_\mu(E_\nu^{\rm obs},E_{\rm thr}),
\end{equation}
where $P_{\rm surv}$ is the probability that a neutrino reaches
the detector without being absorbed by the earth. This is a function of
the neutrino-nucleon interaction length in the earth 
given by $L_{\rm int} =  \{\sigma^{\rm tot}(E_\nu^{\rm obs}) N_A\}^{-1}$,
($N_A$ is the Avogadro number and $\sigma^{\rm tot}$ is the sum total 
of CC and NC interactions) and the effective
path length $X(\theta_z)$ (gm cm$^{-2}$) for incident neutrino
zenith angle $\theta_z$. $P_{\rm surv}$ takes the form, 
$$
P_{\rm surv} (E_\nu^{\rm obs},\theta_z) = \exp [-X(\theta_z)/L_{\rm int}]  
= \exp [-X(\theta_z) \sigma^{\rm tot} N_A ].
$$
In Eq. (16), 
\begin{equation}
P_\mu(E_\nu^{\rm obs},E_{\rm thr}) = N_A \sigma^{\rm CC} 
\langle R(E_\nu^{\rm obs};E_{\rm thr})
\rangle \,\, ,
\end{equation}
with $N_A$ is the Avogadro number and $\sigma^{\rm CC}$
is the $\nu_\mu$ CC interaction. The average range of muon 
inside the rock is given by 
\begin{equation}
\langle R(E_\nu^{\rm obs};E_{\rm thr}) \rangle = \frac {1} {\sigma^{\rm CC}}
\displaystyle\int_0^{1 - E_{\rm thr}/E_\nu} 
dy R(E_\nu^{\rm obs} (1 - y), E_{\rm thr})
\frac {d\sigma^{\rm CC}(E_\nu^{\rm obs},y)} {dy}
\end{equation}
where $y = (E_\nu^{\rm obs} - E_\mu)/E_\nu^{\rm obs}$.
The range $R (E_\mu, E_{\rm thr})$ for a muon
of energy $E_\mu$ is given as
\begin{equation}
R (E_\mu, E_{\rm thr}) = \displaystyle\int^{E_\mu}_{E_{\rm thr}} 
\frac {dE_\mu} {\langle dE_\mu/dX \rangle} \simeq \frac {1} {\beta}
\ln \left ( \frac {\alpha + \beta E_\mu} {\alpha + \beta E_{\rm thr}}.
\right )
\end{equation}
The average lepton energy loss with energy $E_\mu$ per unit distance
travelled is given by
\cite{gaisser}
\begin{equation}
\left \langle \frac {dE_\mu} {dX} \right\rangle = -\alpha - \beta E_\mu
\end{equation}
The values of $\alpha$ and $\beta$ used in the present calculations
are
\begin{eqnarray}
\alpha &=& \{ 2.033 + 0.077\ln[E_\mu {\rm (GeV)}] \}\times 10^{-3} {\rm GeV}
{\rm cm}^2 {\rm gm}^{-1} \nonumber \\
\beta &=& \{ 2.033 + 0.077\ln[E_\mu {\rm (GeV)}] \} \times 10^{-6}
{\rm cm}^2 {\rm gm}^{-1} 
\end{eqnarray}
for $E_\mu \la 10^6$ GeV \cite{dar} and
\begin{eqnarray}
\alpha &=& 2.033 \times 10^{-3} {\rm GeV} 
{\rm cm}^2 {\rm gm}^{-1} \nonumber \\
\beta &=& 3.9 \times 10^{-6} 
{\rm cm}^2 {\rm gm}^{-1}
\end{eqnarray}
otherwise \cite{guetta1}.
 
\section{Calculations and Results}

We define a ratio $R$ of the track events and shower events at the kilometer 
scale detector considered here.
\begin{equation}
R = \frac {\rm Muon\,\,\,track\,\,\,events} {\rm Shower\,\,events}
\end{equation}
The muon track events are therefore the total sum of the events 
from $\nu_\mu$ CC interactions and $\nu_\tau$ CC interactions 
(as described in the previous section). They are obtained 
using Eqs (16 - 22) with the flux in the expression for $S$ (total 
number of secondary muons) of Eq. (16) suitably replaced by $\phi_{\nu_\mu}$ 
or  $\phi_{\nu_\tau}$ (Eq. 15) as the case may be. The neutrino 
CC and NC cross-sections ($\sigma^{CC}$ and $\sigma^{NC}$) are taken from 
Ref. \cite{raj2}. The effective path length used in the expression 
for $P_{\rm surv}$ (Eq. (16)) is written as 
$X(\theta_z) = \int \rho (r(\theta_z, \ell) d\ell$
where $\rho (r(\theta_z, \ell)$ is the matter density inside the earth
at a distance
$r$ from the centre of the earth and $\ell$ is the neutrino path 
length with zenith angle $\theta_z$. The earth matter density is taken from 
Ref. \cite{raj1} that follows from preliminary earth reference model 
(PREM). The calculations are made for GRB neutrino zenith angle 
$\theta_z = 100.9^o$. 

The shower 
events are obtained using the equation 
\begin{equation}
N_{\rm sh} = \int dE_\nu \frac {dN_{\nu}} {dE_\nu} P_{\rm surv}(E_\nu)
\times \int \frac {1} {\sigma^j} \frac {d {\sigma^j}} {dy} P_{\rm int}
(E_\nu,y)\,\,,
\end{equation} 
where, $\sigma^j = \sigma^{\rm CC}$ (for electromagnetic shower
from $\nu_e$ charged current interactions) or $\sigma^{\rm NC}$ as the
case may be and $P_{\rm int}$ is the probability that a
shower produced by the neutrino interactions will be detected and is given by
\begin{equation}
P_{\rm sh} = \rho N_A \sigma^j L
\end{equation}
where $\rho$ is the density of the detector material and $L$ is the
length of the detector (L = 1 Km for IceCube). Here too, the flux 
$dN_{\nu}/ {dE_\nu}$ in Eq. (24) is to be replaced by 
$\phi_{\nu_\mu}$,  $\phi_{\nu_\tau}$ or $\phi_{\nu_e}$ (Eq. 15) 
as the case may be. 
 
From Eqs. (8,9), we see that the unparticle decay of neutrinos 
depend on the scaling dimension $d_{\cal U}$ (non integral number) and the 
neutrino-unparticle coupling strength $\lambda_\nu^{ij}$. In what 
follows, both $\lambda_\nu^{ij}$ and just $\lambda$ both will signify 
the same coupling.   
In order study how a possible unparticle decay of neutrinos can 
affect the UHE neutrino signal from GRB, we vary these parameters 
and the subsequent variation of the ratio $R$ (Eq. 23) is calculated 
with the formalism discussed so far. The unparticle decay of neutrinos 
considered 
here, $|\nu_2 \rangle$ and $|\nu_3 \rangle$ are considered unstable and 
subject to undergo unparticle decay while only 
$|\nu_1 \rangle$ is stable. Therefore, in Eq. (8) we need the masses $m_2$ and 
$m_3$ for unparticle decay of respective neutrinos. In the present
calculations the value of $m_2$ is estimated from 
$m_2 = \sqrt {\Delta m_{32}^2}$, where 
$\Delta m_{32}^2 = m_3^2 - m_2^2$ (normal hierarchy) 
and $\Delta m_{32}^2= 2.5 \times 10^{-3}$ eV$^2$  
(from atmospheric neutrino oscillation) for the present work. 
The value of $m_3$ thus follows. 
The cutoff scale $\Lambda_{\cal U}$ is taken to be 1 TeV.   

Fig. 1 shows the variation of $R$ with coupling strength $\lambda_\nu^{ij}$
for three different values of scaling dimension  $d_{\cal U}$ for a neutrinos
from a GRB with fixed redshift $z = 0.03$. This redshift corresponds to 
a baseline length of $3.8 \times 10^{21}$ Km. The ratio 
$R$ becomes insensitive to the variation of coupling for higher values 
of $d_{\cal U}$. This phenomenon can be understood from Fig. 2 where 
the variation of decay lifetime $\tau$ (in terms of $\frac {\tau} {m}$)
with $d_{\cal U}$ is shown. This plot clearly shows that $\frac {\tau} {m}$
increases with the increase of $d_{\cal U}$. From Fig. 2, for example,
$\frac {\tau} {m} \sim 10^{18}$ when  $d_{\cal U} = 1.3$. 
For UHE neutrinos with an energy 
$\sim 10^6$ GeV coming form a GRB at 100 Megaparsec distance 
($L \sim 10^{21}$ Km), exponential decay term in Eq. (10) tends to 1 and 
thus the muon track to shower ratio ($R$) tends to that for 
mass flavour oscillation. 
The value of the ratio $R$ in case of mass flavour oscillation scenario
for the same set of oscillation parameters and baseline length of 
GRB neutrinos ($z = 0.03$) 
is calculated as $R_{\rm mass-flavour} = 3.23$ 
and for no oscillation, $R$ is computed as  $R_{\rm no\,osc} = 5.6$ for 
the same GRB. 

Since the coupling $\lambda_\nu^{ij}$ 
appears at the denominator of the expression for 
$\frac {\tau} {m}$ (Eq. 8), for higher values of coupling the decay effects 
may appear if for those values, the decay length $\lambda_d$ becomes 
$\sim$ the baseline length $L$. This is also apparent in Fig. 1. 
One also observes from Fig. 1 that for 
suitable values of scale dimension $d_{\cal U}$ and the unparticle coupling 
srength $\lambda$, the ratio $R$ can differ from mass flavour value 
(without the decay) and also from no oscillation value to a considerable 
extent (in fact the value $R$ can even be
very close to 0). Therefore, for certain conditions of unparticle 
scenario, the unparticle decay of neutrinos may indeed be probed 
by observing the UHE signal at a kilometer scale detector like IceCube. 

In Fig. 3, the variations of the muon track to shower ratio $R$ with 
$d_{\cal U}$ are shown for four different values (0.0001, 0.001, 0.01 
and 0.1) of couplings 
$\lambda^{ij}_\nu$. For higher values of $d_{\cal U}$ and for 
lower values of $\lambda$, decay effects vanish and the value of $R$ tends to 
that for only mass-flavour oscillation. A fact that can be understood 
from Eqs. 8-11 and also discussed above. But once again, Fig. 3 shows 
substantial varitaion of the results from those obtained 
considering only mass-flavour oscillation, 
for certain ranges of values of $d_{\cal U}$ and coupling 
$\lambda^{ij}_\nu$. For example, for $d_{\cal U} = 1.2$ and 
$\lambda = 0.001$, the value of $R \simeq 1.27$ $--$ a variation of 
more than 60\% from the mass-flavour oscillation value. Again, to obtain
similar effects for higher value of $d_{\cal U}$, the value of $\lambda$
is also to be increased. But perhaps, the unparticle coupling with 
neutrino may not be very high and the scale dimension $d_{\cal U}$ is to 
remain within the value $1 < d_{\cal U} < 2$. This may be mentioned 
that in Ref. \cite{chenHe2} the unparticle decay of neutrinos is used to 
study the constraints on $d_{\cal U}$, $\Lambda_{\cal U}$ etc. 

The muon yield and muon to shower ratio are also affected by the 
GRB neutrino flux. The GRB neutrino flux has a $d^{-2}(z)$ dependence,
(Eq. 4) where $d(z)$ is the distance of a GRB with redshift $z$. The 
flux also ofcourse has a $E^{-2}$ dependence (Eq. 1). In an
event of the unparticle decay of neutrinos, the variation of muon to shower 
ratio ($R$) signal at a kilometer scale detector like IceCube 
for different GRBs at different redshift distances 
is investigated in Fig. 4. The variations of $R$ with $z$ are plotted 
for three different values of unparticle coupling 
($\lambda = 0.0001, 0.001, 0.01$). Similar variation for pure mass-flavour 
oscillation and for no oscillation or decay are also shown in Fig. 4.
From Fig. 4, one sees that to obtain 
a variation of $\sim 50\% - 60\%$ for the value of $R$ from that of 
the pure mass-flavour oscillation value, the coupling is to be large 
or redshift $z$ (and hence GRB distance $d(z)$) also should be large enough.
For the latter case, however, the neutrino flux from GRB will be 
reduced (Eq. 4) which may affect significant detection. 

\section{Summary and Discussions}

We have considered here the recently proposed ``unparticle stuff" associated
with the possible existence of a nontrivial scale invariant sector 
with scale dimension $d_{\cal U}$. Such ``unparticle stuff" appears 
to be a nonintegral number $d_{\cal U}$ of invisible particles. 
Unparticles may open up unexpected phenomenology and in the present 
work we consider the decay of neutrinos to unparticles. The consequences
of such decay process is explored with the example of ultra high 
energy neutrinos from cosmological Gamma Ray Bursts. We calculate
how the unparticle decay of neutrinos, if exists, affects the 
neutrino fluxes (of all active flavours) from GRB on arriving at earth. 
We then attempt to estimate how the effect a possible unparticle decay of 
neutrinos can change the detection yield of GRB neutrinos from 
only mass-flavour oscillation (or no oscillation) at
a kilometer scale ice Cerenkov detector like IceCube. The GRB 
neutrinos are chosen because of the astronomical baselines of 
GRB neutrinos (for a terrestrial detector), the oscillatory part 
is averaged out and one obtains an overall suppression for 
muon neutrinos. For the present calculation we probe the ratio 
$R$ of the total secondary muon tracks to the total shower events
at the detector for various possible scenarios. The muons and the 
showers are obtained as the secondary particles following the 
CC and NC interactions of the cosmic neutrinos with the terrestrial 
rock and detector material.  
We obtain more than $60\% - 65\%$ deviation of the value of $R$ from 
only the mass-flavour oscillation scenario for 
some cases of unparticle decay of neutrinos considered here.

Such deviations can be 
probed by an  efficient detector. The volume of IceCube is  
1 Km$^3$ but this is the volume over which the optical modules 
(OM) are sparsed. Even if a CC interaction vertex occurs in the ice 
outside this volume, the resulting secondary particle can be detected 
by IceCube if it enters the specified detector volume. Thus the effective
volume of the detector is in fact more than 1 Km$^3$. Although, IceCube
detector has better energy and angular resolutions, there are backgrounds
from cosmic muons and atmospheric neutrinos. Simulation study of 
IceCube detector by Ahrens et al \cite{ahrens} indicates that 
the cosmic neutrino signal 
from the diffuse cosmic neutrino flux is well below the atmospheric 
limit. A diffuse cosmic isotropic flux results from the summation 
of the cosmic sources. 
But here, we are considering the flux from specific GRBs.
Thus they are like point sources. In this case, one expects an 
excess of events from a particular directions. This considerably 
reduces the background if the detector has very good angular resolution. 
The simulation done at Ref. \cite{ahrens} with 1$^o$ angular search 
cone shows an improvement of the measurement with the diffuse flux.       
Moreover, the atmospheric neutrino spectrum falls more steeply 
($\sim E^{-3.7}$) as against hard GRB neutrino spectrum ($\sim E^{-2}$) 
from the shock acceleration mechanism considered here. Considering 
all these, a very large deviation of the signal from the expected 
ones (e.g. the ones expected from mass-flavour oscillation) 
may be probed with IceCube. But detailed simulation is required 
to estimate the deviation of the results that can be probed by the 
detector. In the present work, the purpose is to study the effects 
if neutrino decay to unparticles vis-a-vis undergoes mass-flavour oscillation 
and no oscillation and GRB neutrinos are considered for that purpose.

\end{document}